\pdfoutput=1
\RequirePackage{ifpdf}
\ifpdf 
\documentclass[pdftex]{sigma}
\else
\documentclass{sigma}
\fi

\usepackage{braket}
\usepackage[all]{xy}

\newcommand{\p}[1]{\left(#1\right)}

\newcommand{\bb}[1]{\mathbb{#1}}

\newcommand{\app}[3]{#1\colon #2\to #3}

\newcommand{\ol}[1]{\overline{#1}}
\newcommand{\smm}[1]{\left(\begin{smallmatrix} #1\end{smallmatrix}\right)}

\DeclareMathOperator{\id}{Id}
\DeclareMathOperator{\diag}{diag}
\DeclareMathOperator{\tr}{Tr}

\DeclareMathOperator{\GL}{GL}
\DeclareMathOperator{\SL}{SL}

\DeclareMathOperator{\PGL}{PGL}

\DeclareMathOperator{\Mat}{Mat}
\DeclareMathOperator{\Aut}{Aut}

\DeclareMathOperator{\TAut}{TAut}
\DeclareMathOperator{\PTAut}{PTAut}
\DeclareMathOperator{\Aff}{Aff}
\DeclareMathOperator{\Rep}{Rep}
\DeclareMathOperator{\Gr}{Gr}

\begin{document}

\allowdisplaybreaks

\renewcommand{\PaperNumber}{037}

\FirstPageHeading

\ShortArticleName{A Note on the Automorphism Group of the Bielawski--Pidstrygach Quiver}

\ArticleName{A Note on the Automorphism Group\\
of the Bielawski--Pidstrygach Quiver}

\Author{Igor MENCATTINI and Alberto TACCHELLA}
\AuthorNameForHeading{I.~Mencattini and A.~Tacchella}
\Address{ICMC - Universidade de S\~ao Paulo, Avenida Trabalhador S\~ao-carlense, 400,\\
13566-590 S\~ao Carlos - SP, Brasil}
\Email{\href{mailto:igorre@icmc.usp.br}{igorre@icmc.usp.br}, \href{mailto:tacchella@icmc.usp.br}{tacchella@icmc.usp.br}}

\ArticleDates{Received August 29, 2012, in f\/inal form April 26, 2013; Published online April 30, 2013}

\Abstract{We show that there exists a~morphism between a~group $\Gamma^{\mathrm{alg}}$ introduced by
G.~Wilson and a~quotient of the group of tame symplectic automorphisms of the path algebra of a~quiver
introduced by Bielawski and Pidstrygach.
The latter is known to act transitively on the phase space \(\mathcal{C}_{n,2}\) of the Gibbons--Hermsen
integrable system of rank $2$, and we prove that the subgroup generated by the image of
$\Gamma^{\mathrm{alg}}$ together with a~particular tame symplectic automorphism has the property that, for
every pair of points of the regular and semisimple locus of \(\mathcal{C}_{n,2}\), the subgroup contains an
element sending the f\/irst point to the second.}


\Classification{37K10; 16G20; 14A22}

\section{Introduction}

Let \(n\) and \(r\) be two positive natural numbers and denote by \(\Mat_{n,r}(\bb{C})\) the complex vector
space of \(n\times r\) matrices with entries in \(\bb{C}\).
The space
\begin{gather*}
V_{n,r}\mathrel{:=}\Mat_{n,n}(\bb{C})\oplus\Mat_{n,n}(\bb{C})\oplus\Mat_{n,r}(\bb{C})\oplus\Mat_{r,n}(\bb{C})
\end{gather*}
can be viewed (using the identif\/ications provided by the trace form) as the cotangent bundle of the
vector space \(\Mat_{n,n}(\bb{C})\oplus \Mat_{n,r}(\bb{C})\), thus it comes equipped with the canonical
holomorphic symplectic form
\begin{gather}
\omega(X,Y,v,w)=\tr(\mathrm{d}X\wedge\mathrm{d}Y+\mathrm{d}v\wedge\mathrm{d}w).
\label{eq:ss}
\end{gather}
The group \(\GL_{n}(\bb{C})\) acts on \(V_{n,r}\) by
\begin{gather}
\label{eq:az}
g.(X,Y,v,w)=\big(gXg^{-1},gYg^{-1},gv,wg^{-1}\big).
\end{gather}
This action is Hamiltonian, and the corresponding moment map
\(\app{\mu}{V_{n,r}}{\mathfrak{gl}_{n}(\bb{C})}\) is
\begin{gather}
\label{eq:mom-mu}
\mu(X,Y,v,w)=[X,Y]+vw.
\end{gather}
For every complex number \(\tau\neq 0\) the action of \(\GL_{n}(\bb{C})\) on \(\mu^{-1}(\tau I)\) is free,
hence we can perform the symplectic quotient
\begin{gather}
\label{eq:Cnr}
\mathcal{C}_{n,r}\mathrel{:=}\mu^{-1}(\tau I)/\GL_{n}(\bb{C}).
\end{gather}
This family of smooth, irreducible af\/f\/ine algebraic varieties plays an important r\^ole in various
f\/ields.
They are examples of Nakajima quiver varieties~\cite{naka94}, and they also arise in the work of Nekrasov
and Schwarz~\cite{ns98} on the moduli space of instantons on a~non-commutative \(\bb{R}^{4}\).
Finally, and most importantly from the perspective of the present work, they can be seen as a~completion of
the phase space of a~family of integrable systems that generalize the well-known rational Calogero--Moser
model.

\subsection[The Gibbons-Hermsen system]{The Gibbons--Hermsen system}

Let us brief\/ly remind the reader of the def\/inition of this integrable system, as it was introduced by
Gibbons and Hermsen in the paper~\cite{gh84}.
Just like the (complexif\/ied) Calogero--Moser model, the system describes the motion of $n$ point
particles in the complex plane interacting pairwise according to a~potential proportional to the second
inverse power of their distance.
In addition to the Calogero--Moser case, however, each particle is endowed with some additional
``internal'' degrees of freedom, parametrized by a~vector \(v_{i}\) in an auxiliary vector space
$V\simeq\bb{C}^r$ and by its canonical conjugate \(\xi_{i}\) in the dual space $V^{\ast}$.
The Hamiltonian of the system is given by
\begin{gather}
\label{eq:ghH}
H(x,p,v,\xi)=\frac{1}{2}\sum_{i=1}^{n}p_{i}^{2}+\frac{1}{2}\sum_{i,j=1}^{n}\frac{\xi_i(v_j)\xi_j(v_i)}
{(x_i-x_j)^{2}}.
\end{gather}
For each particle \(i\) there is the constraint $\xi_{i}(v_{i})=-1$ (notice that these quantities are
constants of the motion); moreover, two pairs \((v_{i},\xi_{i})\) and \((v'_{i},\xi'_{i})\) are considered
equivalent if \(v_{i} = \lambda v'_{i}\) and \(\xi_{i} = \lambda^{-1}\xi'_{i}\) for some \(\lambda\in
\bb{C}^{*}\).
When \(r=1\) these requirements completely f\/ix the additional degrees of freedom and we recover the
classic rational Calogero--Moser system.

As it was proved by Gibbons and Hermsen, the Hamiltonian system described above is completely integrable
and its phase space can be identif\/ied with the manifold \(\mathcal{C}_{n,r}\).
Let us explain shortly how goes the proof of the complete integrability.
Consider, for each \(k\in \bb{N}\) and \(\alpha\in \mathfrak{gl}_{r}(\bb{C})\), the following function on
the space \(V_{n,r}\):
\begin{gather}
\label{eq:fh}
J_{k,\alpha}=\tr Y^{k}v\alpha w.
\end{gather}
These functions are invariant with respect to the action~\eqref{eq:az}, so that they descend to well
def\/ined functions on the quotient space \(\mathcal{C}_{n,r}\); the Hamiltonian~\eqref{eq:ghH} coincides,
up to scalar multiples, with~$J_{2,I}$.
The equations of motion determined by \(J_{k,\alpha}\) are
\begin{gather*}
\dot{X}=Y^{k-1}v\alpha w+Y^{k-2}v\alpha wY+\cdots +v\alpha wY^{k-1},
\\
\dot{Y}=0,
\\
\dot{v}=Y^kv\alpha,
\\
\dot{w}=-\alpha wY^{k},
\end{gather*}
where $\dot{A}=\frac{dA}{dt}$.
From this we can deduce that the Gibbons--Hermsen f\/lows are complete.
In fact, since $Y$ is constant, the equations for $v$ and $w$ are linear with constant coef\/f\/icients.
This implies that the solutions of the last two equations are linear combinations of polynomials and
exponentials, forcing the solution of the f\/irst equation to be of the same form.

The Poisson brackets def\/ined by the symplectic form~\eqref{eq:ss} are given by
\begin{gather*}
\{X_{ij},Y_{k\ell}\}=\delta_{jk}\delta_{i\ell}
\qquad
\text{and}
\qquad
\{v_{ij},w_{k\ell}\}=\delta_{jk}\delta_{i\ell},
\end{gather*}
all the others being equal to zero.
Then a~short calculation shows that the Poisson bracket between two functions of the form~\eqref{eq:fh} is
\begin{gather}
\label{eq:pbh}
\{J_{m,\alpha},J_{\ell,\beta}\}=J_{m+\ell,[\alpha,\beta]},
\end{gather}
where $[\cdot,\cdot]$ is the matrix commutator.
Notice that these are the same relations holding in the Lie algebra of polynomial loops in
$\mathfrak{gl}_r(\bb{C})$: explicitly, the correspondence is given by
\begin{gather}
\label{eq:corr}
J_{k,\alpha}\leftrightarrow z^{k}\alpha.
\end{gather}
From~\eqref{eq:pbh} follows in particular that \(\{J_{m,\alpha}, J_{\ell,\beta}\} = 0\) if and only if
\([\alpha,\beta] = 0\).
It is then possible to f\/ind among the functions~\eqref{eq:fh} a~total of \(nr\) independent and mutually
commuting f\/irst integrals (e.g.\
by taking \(1\leq k\leq n\) and matrices \(\alpha\) spanning the space of diagonal \(r\times r\) matrices).
These results imply the complete integrability of the Gibbons--Hermsen system.

\subsection{Some reminder about non-commutative symplectic geometry}

In~\cite{bp08}, Bielawski and Pidstrygach study the varieties~\eqref{eq:Cnr} in the case \(r=2\) using the methods
of non-commutative symplectic geometry~\cite{blb02, ginz01}, starting from the quiver
\begin{gather}
\label{eq:quiv-bp}
Q_{\mathrm{BP}}=\xymatrix{\bullet_{1}\ar@(ul,dl)[]_{a}\ar@/_/[r]_{y}&\bullet_{2}\ar@/_/[l]_{x}}.
\end{gather}
Recall that a~quiver is simply a~directed graph, possibly with loops and multiple edges.
To every quiver \(Q\) one can associate its \emph{double} \(\ol{Q}\) obtained by keeping the same vertices
and adding, for each arrow \(\app{\rho}{i}{j}\), a~corresponding arrow \(\app{\rho^{*}}{j}{i}\) going in
the opposite direction.
The \emph{path algebra} (over \(\bb{C}\)) of a~quiver \(Q\), denoted \(\bb{C}Q\), is the complex
associative algebra which is generated, as a~linear space, by all the paths in \(Q\) and whose product is
given by composition of paths (or zero when two paths do not compose).

Now let \(A\) denote \(\bb{C}\ol{Q}_{\mathrm{BP}}\), the path algebra of the double of \(Q_{\mathrm{BP}}\).
Denote also by \(\TAut(A;c)\) the group of (tame) \emph{non-commutative symplectomorphisms} of this algebra
(see Def\/inition~\ref{def:TAut} in the next section).
One of the main results of~\cite{bp08} is that this group acts \emph{transitively} on \(\mathcal{C}_{n,2}\).
This is to be compared with the well-known result for the case \(r=1\), f\/irst obtained by Berest and
Wilson in~\cite{bw00}, according to which the group \(G_{\mathrm{CM}}\) of automorphisms of the f\/irst
Weyl algebra
\begin{gather*}
A_{1}=\bb{C}\langle a,a^{*}\rangle/(aa^{*}-a^{*}a-1)
\end{gather*}
acts transitively on the Calogero--Moser varieties \(\mathcal{C}_{n,1}\).
This group can also be interpreted from the perspective of non-commutative symplectic geometry in the
following way.
Let \(Q_{\circ}\) denote the quiver with one vertex and one loop \(a\) on it.
The path algebra of its double \(\ol{Q}_{\circ}\) is just the free associative algebra on the two
generators \(a\) and \(a^{*}\).
The group of non-commutative symplectomorphisms of this algebra is the group of automorphisms of
\(\bb{C}\langle a,a^{*}\rangle\) preserving the commutator \([a,a^{*}]\), and this group is isomorphic to
\(G_{\mathrm{CM}}\) by a~result of Makar-Limanov~\cite{ml84, ml70}.
Hence the rank~1 case f\/its into the same picture, by replacing the quiver~\eqref{eq:quiv-bp} with
\(Q_{\circ}\).

It turns out that in this case these non-commutative symplectomorphisms have a~very natural interpretation
in terms of f\/lows of the Calogero--Moser system.
Indeed, a~classic result of Dixmier~\cite{dix68} implies that the group \(G_{\mathrm{CM}}\) is generated by
a~family of automorphisms \(\Phi_{p}\) labeled by a~polynomial \(p\) in \(a^{*}\) (say with zero constant
term), def\/ined by the following action on the generators of \(A_{1}\):
\begin{gather}
\label{eq:def-cmflows}
\Phi_{p}(a)=a-p'\big(a^{*}\big),
\qquad
\Phi_{p}\big(a^{*}\big)=a^{*},
\end{gather}
together with the single automorphism \(\mathcal{F}_{0}\) def\/ined by
\begin{gather}
\label{eq:fft0}
\mathcal{F}_{0}\big(a,a^{*}\big)=(-a^{*},a),
\end{gather}
that we will call the \emph{formal Fourier transform}.
The action of these generators of \(G_{\mathrm{CM}}\) on \(\mathcal{C}_{n,1}\) is given by
\begin{gather*}
\Phi_{p}.(X,Y,v,w)=(X-p'(Y),Y,v,w),
\qquad
\mathcal{F}_{0}.(X,Y,v,w)=(-Y,X,v,w).
\end{gather*}
In particular the action of \(\Phi_{p}\) for a~given polynomial \(p = p_{1}a^{*} + p_{2}a^{*2} + \cdots\)
corresponds exactly to the action of a~linear combination of the (mutually commuting) Calogero--Moser
f\/lows, i.e.\
the f\/lows of the Hamiltonian functions \(\tr Y^{k}\) (for \(k\geq 1\)) on \(\mathcal{C}_{n,1}\), with
``times'' \((p_{1}, p_{2}, \dots)\).

\subsection{The main results of this note}

Given the above, it is natural to ask if a~similar picture holds also in the rank 2 case; namely, if the
action of the group \(\TAut(A;c)\) considered by Bielawski and Pidstrygach on \(\mathcal{C}_{n,2}\) can be
made more concrete by interpreting its elements as f\/lows of the Gibbons--Hermsen
Hamiltonians~\eqref{eq:fh}.
One dif\/f\/iculty here is given by the fact that, while the Calogero--Moser Hamiltonians generate an
abelian Lie algebra, the Hamiltonians~\eqref{eq:fh} generate a~non-abelian one that cannot be trivially
exponentiated to get a~Lie group.
In other words, when \(r>1\) the maps of the form \(\exp p(z)\), where \(p\) is a~polynomial map
\(\bb{C}\to \mathfrak{gl}_{r}(\bb{C})\), do not form a~group.
One way to avoid this problem would be to simply take the group \(\Gamma\) of \emph{all} holomorphic maps
\(\bb{C}\to \GL_{2}(\bb{C})\), but this group contains elements giving non-polynomial f\/lows on
\(\mathcal{C}_{n,2}\) which cannot be realized by the action of an element in \(\TAut(A;c)\).

In the unpublished notes~\cite{wils09}, G.~Wilson suggests to consider instead the subgroup of \(\Gamma\)
def\/ined by
\begin{gather}
\label{eq:def-Galg}
\Gamma^{\mathrm{alg}}\mathrel{:=}\Gamma^{\mathrm{alg}}_{\text{sc}}\times\PGL_{2}(\bb{C}[z]),
\end{gather}
where \(\Gamma^{\mathrm{alg}}_{\text{sc}}\) is the subgroup consisting of maps of the form \(\mathrm{e}^{p}
I_{2}\) for some polynomial \(p\) with no constant term and \(\PGL_{2}(\bb{C}[z])\) is seen as a~subgroup
of \(\Gamma\) in the obvious manner.
(The choice of this particular subgroup can be motivated also on purely algebraic grounds, as we will
explain in Section~\ref{concl}.)

Now denote by \(\PTAut(A;c)\) the quotient of \(\TAut(A;c)\) by the subgroup of \emph{scalar} af\/f\/ine
symplectic automorphisms, whose action on \(\mathcal{C}_{n,2}\) is trivial (see
Def\/inition~\ref{def:PTAut} below).
The aim of this note is to prove the following:
\begin{theorem}
\label{teo:result}
There exists a~morphism of groups
\begin{gather}
\label{eq:map}
\app{i}{\Gamma^{\mathrm{alg}}}{\PTAut(A;c)},
\end{gather}
such that, if \(\mathcal{P}\) denotes the subgroup of \(\PTAut(A;c)\) generated by the image of \(i\) and
the symplectomorphism \(\mathcal{F}\) defined by extending the automorphism~\eqref{eq:fft0} from
\(\bb{C}\ol{Q}_{\circ}\) to \(\bb{C}\ol{Q}_{\mathrm{BP}}\) in the following way:
\begin{gather}
\label{eq:def-fft}
\mathcal{F}\big(a,a^{*},x,x^{*},y,y^{*}\big)\mathrel{:=}(-a^{*},a,-y^{*},y,-x^{*},x)
\end{gather}
and \(\mathcal{R}_{n,2}\) is the Zariski open subset of \(\mathcal{C}_{n,2}\) consisting of quadruples
\(\p{X,Y,v,w}\) such that either~\(X\) or~\(Y\) is regular semisimple $($i.e., diagonalizable with distinct
eigenvalues$)$, then for every pair of points \(\xi_1,\xi_2\in \mathcal{R}_{n,2}\) there exists an element of
\(\mathcal{P}\) which maps \(\xi_1\) to \(\xi_2\).
\end{theorem}

Here \(\mathcal{R}_{n,2}\) should be seen as the rank~2 version of the analogue subset of the
Calogero--Moser space consisting of quadruples \((X,Y,v,w)\) for which either \(X\) or \(Y\) are
diagonalizable.

\begin{remark}
Theorem~\ref{teo:result} is not a~real transitivity result since the action of \(\mathcal{P}\) does not
preserve the subset \(\mathcal{R}_{n,2}\).
Unfortunately, it is not easy to understand if this action is transitive on the whole of
\(\mathcal{C}_{n,2}\) or not.
The main dif\/f\/iculty comes from the fact that the proof of transitivity in~\cite{bp08} for points
outside of \(\mathcal{R}_{n,2}\) is not constructive; for this reason studying the action of
\(\PTAut(A;c)\) on such points is much more dif\/f\/icult.
\end{remark}

\section{Preliminaries}

For the remainder of this paper, \(r\) will be f\/ixed and equal to~\(2\).
In this case, as noticed in~\cite{bp08}, we can obtain the manifold \(\mathcal{C}_{n,2}\) def\/ined
by~\eqref{eq:Cnr} starting from the space of representations of the double \(\ol{Q}_{\mathrm{BP}}\) of the
quiver~\eqref{eq:quiv-bp}, in the following manner.

\looseness=-1
Let us denote by \(\Rep(\ol{Q}_{\mathrm{BP}},(n,1))\) the complex vector space of linear representations of
\(\ol{Q}_{\mathrm{BP}}\) with dimension vector \((n,1)\).
A point in this space is a~\(6\)-tuple \((A,B,X_{1},X_{2},Y_{1},Y_{2})\) consisting of two \(n\times n\)
matrices, two \(n\times 1\) matrices and two \(1\times n\) matrices that represent, respectively, the
arrows \(a\), \(a^{*}\), \(x\), \(y^{*}\), \(y\) and \(x^{*}\) in \(\ol{Q}_{\mathrm{BP}}\).
This space is in bijection with \(V_{n,2}\) via the following map:
\begin{gather}
\label{eq:m1}
A\mapsto X,
\qquad
B\mapsto Y,
\qquad
X_{1}\mapsto v_{\bullet1},
\qquad
X_{2}\mapsto-v_{\bullet2},
\qquad
Y_{1}\mapsto w_{2\bullet},
\qquad
Y_{2}\mapsto w_{1\bullet},
\end{gather}
where by \(v_{\bullet i}\) we denote the \(i\)-th column of the \(n\times 2\) matrix \(v\), and similarly
by \(w_{j\bullet}\) we denote the \(j\)-th row of the \(2\times n\) matrix \(w\).

On the space \(\Rep(\ol{Q}_{\mathrm{BP}},(n,1))\) there is a~natural action of the group
\begin{gather*}
G_{(n,1)}=(\GL_{n}(\bb{C})\times\GL_{1}(\bb{C}))/\bb{C}^{*}\simeq\GL_{n}(\bb{C})
\end{gather*}
(where \(\bb{C}^{*}\) is seen as the subgroup of pairs of the form \((\lambda I_{n},\lambda)\) for some
\(\lambda\in \bb{C}^{*}\)) by change of basis.
This action is Hamiltonian, with moment map given by
\begin{gather}
\label{eq:mom-nu}
\nu(A,B,X_{1},X_{2},Y_{1},Y_{2})=([A,B]+X_{1}Y_{2}-X_{2}Y_{1},Y_{1}X_{2}-Y_{2}X_{1})\in\mathfrak{g}_{(n,1)}.
\end{gather}
It is easy to verify that, under the bijection~\eqref{eq:m1}, this action of \(G_{(n,1)}\) on
\(\Rep(\ol{Q}_{\mathrm{BP}},(n,1))\) precisely coincides with the action of \(\GL_{n}(\bb{C})\) on
\(V_{n,2}\) given by~\eqref{eq:az}.
Finally, by comparing the two moment maps~\eqref{eq:mom-mu} and~\eqref{eq:mom-nu}, we conclude that
\(\mathcal{C}_{n,2}\) is exactly the same as the symplectic quotient \(\mu^{-1}(\mathcal{O})/G_{(n,1)}\),
where \(\mathcal{O}\) denotes the coadjoint orbit of the point \((\tau I_{n}, -n\tau)\in
\mathfrak{g}_{(n,1)}\).

As in the introduction, we let \(A\) stand for the path algebra \(\bb{C}\ol{Q}_{\mathrm{BP}}\) of
\(\ol{Q}_{\mathrm{BP}}\); it is a~non-commutative algebra over the ring \(\bb{C}^{2} = \bb{C}e_{1} \oplus
\bb{C}e_{2}\), where the idempotents \(e_{1}\) and \(e_{2}\) correspond to the trivial paths at vertices
\(1\) and \(2\), respectively.
For every \(p\in A\) we denote by \(\Aut (A;p)\) the subgroup of \(\Aut A\) that f\/ixes \(p\).
In particular we will be interested in \(\Aut (A;c)\), where
\begin{gather*}
c\mathrel{:=}[a,a^{*}]+[x,x^{*}]+[y,y^{*}].
\end{gather*}

\begin{definition}
The group \(\Aut (A;c)\) will be called the \textit{group of non-commutative symplectic automorphisms} of
\(A\)~\cite{bp08, kont93}.
\end{definition}

In what follows we will be interested in the following types of elements of \(\Aut (A;c)\).
\begin{definition}
An automorphism of \(A\) will be called:
\begin{itemize}\itemsep=0pt
\item \textit{strictly triangular} if it f\/ixes the arrows of \(Q_{\mathrm{BP}}\) (i.e.
\(a\), \(x\) and \(y\));
\item \textit{strictly op-triangular} if it f\/ixes the arrows of
\(Q_{\mathrm{BP}}^{\mathrm{op}}\) (i.e.\
\(a^{*}\), \(x^{*}\), \(y^{*}\)).
\end{itemize}
\end{definition}

An explicit description of strictly triangular symplectic automorphisms of \(\bb{C}\ol{Q}_{\mathrm{BP}}\)
is derived in~\cite{bp08}.
Namely, let~\(F_{2}\) be the free algebra on two generators over~\(\bb{C}\) and def\/ine
\begin{gather}
\label{eq:def-L2}
L_{2}\mathrel{:=}\frac{F_{2}}{\bb{C}+[F_{2},F_{2}]}
\end{gather}
as \looseness=-1 a~quotient of complex vector spaces.
Call \(a\) and \(b\) (the image in \(L_{2}\) of) the two generators of~\(F_{2}\).
Notice that \(L_{2}\) is just the vector space of necklace words in \(a\) and \(b\) (modulo scalars).
Then to every \(f\in L_{2}\) we can associate the automorphism \(\Lambda(f)\in \Aut A\) def\/ined on the
generators of~\(A\) by
\begin{alignat*}{3}
& a\mapsto a, \qquad && a^{*}\mapsto a^{*}+\frac{\partial f}{\partial a}, &\\
& x\mapsto x, \qquad && x^{*}\mapsto x^{*}+y\frac{\partial f}{\partial b}, & \\
& y\mapsto y, \qquad && y^{*}\mapsto y^{*}+\frac{\partial f}{\partial b}x, &
\end{alignat*}
where the substitution \(b=xy\) is understood and the \(\bb{C}\)-linear maps
\begin{gather*}
\app{\frac{\partial}{\partial a},\,\frac{\partial}{\partial b}}{L_{2}}{F_{2}}
\end{gather*}
are the ``necklace derivations'' def\/ined e.g.\ in~\cite{blb02, ginz01}.
Explicitly, they act as usual derivations, except that the letters in a~necklace word must be cyclically
permuted in order to always bring the cancelled letter at the front.

\begin{example}
Let $f_1=aab$ and $f_2=aaab$.
Then
\begin{gather*}
\frac{\partial}{\partial a}f_1=ab+ba,\qquad \text{and}\qquad \frac{\partial}{\partial a}f_2=aab+aba+baa.
\end{gather*}
More generally
\begin{gather*}
\frac{\partial}{\partial a}(a^{n}b)=a^{n-1}b+a^{n-2}ba+\dots+aba^{n-2}+ba^{n-1}.
\end{gather*}
\end{example}

Notice that the result lives in \(F_{2}\), not in \(L_{2}\); in particular it is not a~necklace word, but
a~genuine word in the generators.
\begin{theorem}[Proposition~7.2 in~\cite{bp08}]
\label{teo:bp}
Every \(\Lambda(f)\) is symplectic, and every symplectic automorphism that fixes \(a\), \(x\) and \(y\)
lies in the image of \(\Lambda\).
\end{theorem}

A completely analogous description holds for strictly op-triangular symplectic automorphisms.
Indeed, let \(L_{2}^{\mathrm{op}}\) denote the same vector space~\eqref{eq:def-L2}, but call now \(a^{*}\)
and \(b^{*}\) (the image of) the two generators of \(F_{2}\).
For every \(f\in L_{2}^{\mathrm{op}}\), let \(\Lambda'(f)\) be the strictly op-triangular automorphism of~\(A\) def\/ined by
\begin{alignat*}{3}
& a\mapsto a+\frac{\partial f}{\partial a^{*}}, \qquad && a^{*}\mapsto a^{*}, & \\
& x\mapsto x+\frac{\partial f}{\partial b^{*}}y^{*}, \qquad && x^{*}\mapsto x^{*}, & \\
& y\mapsto y+x^{*}\frac{\partial f}{\partial b^{*}}, \qquad && y^{*}\mapsto y^{*}, &
\end{alignat*}
where \(b^{*} = y^{*}x^{*}\).
We claim that every \(\Lambda'(f)\) is symplectic, and every symplectic automorphism of \(A\) that f\/ixes
\(a^{*}\), \(x^{*}\) and \(y^{*}\) is of this form.
This can easily be proved by recycling exactly the same arguments used in~\cite{bp08} to prove
Theorem~\ref{teo:bp}.
Alternatively, it is easy to verify that an automorphism \(\varphi\) is strictly triangular if and only if
the automorphism \(\mathcal{F}\circ \varphi\circ \mathcal{F}^{-1}\) is strictly op-triangular, where
\(\mathcal{F}\) is the symplectic automorphism def\/ined by~\eqref{eq:def-fft}.
Thus we could simply def\/ine
\begin{gather}
\label{eq:t-opt}
\Lambda'(f\big(a^{*},b^{*}\big))=\mathcal{F}\circ\Lambda(-f(a,b))\circ\mathcal{F}^{-1}.
\end{gather}
Another subgroup of \(\Aut A\) easy to deal with is provided by the \emph{affine} automorphisms, i.e.\
af\/f\/ine trasformations of the linear subspace spanned by \(a\), \(a^{*}\), \(x\), \(x^{*}\), \(y\) and
\(y^{*}\) in \(A\).
An automorphism of this kind which moreover preserves \(c\) is completely specif\/ied by a~pair \((A,T)\)
where \(A\) is an element of \(\bb{C}^{2}\rtimes \SL_{2}(\bb{C})\) (the group of unimodular af\/f\/ine
transformation of \(\bb{C}^{2}\)) acting on the subspace spanned by \(a\) and \(a^{*}\), while \(T\) is an
element of \(\GL_{2}(\bb{C})\) acting as follows on the subspace spanned by the other arrows:
\begin{gather*}
\begin{pmatrix}
-x
\\
y^{*}
\end{pmatrix}
\mapsto T
\begin{pmatrix}
-x
\\
y^{*}
\end{pmatrix},
\qquad
\begin{pmatrix}
x^{*}&y
\end{pmatrix}
\mapsto
\begin{pmatrix}
x^{*}&y
\end{pmatrix}
T^{-1}.
\end{gather*}
Following~\cite{bp08}, we denote by \(\Aff_{c}\) the subgroup consisting of these \emph{affine symplectic
automorphisms}.

\begin{definition}
\label{def:TAut}
The group of \textit{tame symplectic automorphisms of} \(A\), denoted \(\TAut(A;c)\), is the subgroup of
\(\Aut(A;c)\) generated by strictly triangular and af\/f\/ine symplectic automorphisms.
\end{definition}

Notice that the automorphism \(\mathcal{F}\) def\/ined by~\eqref{eq:def-fft} belongs to \(\Aff_{c}\); it
corresponds to the pair determined by \(\smm{0 & -1
\\
1 & 0}\in \SL_{2}(\bb{C})\) and \(\smm{0 & 1
\\
-1 & 0}\in \GL_{2}(\bb{C})\).
It then follows immediately from the relation~\eqref{eq:t-opt} that \(\TAut(A;c)\) can also be generated by
the strictly \emph{op-}triangular automorphisms and by the af\/f\/ine symplectic ones.

Let \(Z\) be the subgroup of \(\TAut(A;c)\) consisting of symplectic af\/f\/ine automorphism of the form
\((I,T)\) where \(I\) is the identity of \(\bb{C}^{2}\rtimes \SL(2,\bb{C})\) and \(T\) belongs to the
center of \(\GL_{2}(\bb{C})\) (i.e.\
\(T = \lambda I\) for some \(\lambda\in \bb{C}^{*}\)).
Then it is easy to see that the action of \(Z\) on \(\mathcal{C}_{n,2}\) is trivial; hence the action of
\(\TAut(A;c)\) on \(\mathcal{C}_{n,2}\) descends through the quotient \(\TAut(A;c)/Z\).

\begin{definition}
\label{def:PTAut}
We denote the quotient \(\TAut(A;c)/Z\) by \(\PTAut(A;c)\).
\end{definition}

An essential r\^ole in the sequel will be played by the following result, f\/irst proved by Nagao
in~\cite{naga59} and later rederived in a~more general context using the Bass--Serre theory of groups acting
on graphs~\cite{serre80}.
Let \(\bb{K}\) be a~f\/ield, and denote by \(B_{2}(\bb{K}[z])\) the subgroup of lower triangular matrices
in \(\GL_{2}(\bb{K}[z])\) and by \(B_{2}(\bb{K})\) the subgroup of lower triangular matrices in
\(\GL_{2}(\bb{K})\).
\begin{theorem}[Nagao]
\label{teo:nagao}
The group \(\GL_{2}(\bb{K}[z])\) coincides with the free product with amalgamation
\(\GL_{2}(\bb{K})\ast_{B_{2}(\bb{K})} B_{2}(\bb{K}[z])\).
\end{theorem}
Suppose now that \(\bb{K}=\bb{C}\).
Then, as is well known, we have that \(B_{2}(\bb{C}) = U_{2}(\bb{C})\rtimes D_{2}(\bb{C})\), where
\(U_{2}(\bb{C})\) is the (normal) subgroup of \emph{lower unitriangular matrices} (= unipotent elements in
\(B_{2}(\bb{C})\)) and \(D_{2}(\bb{C})\) is the subgroup of diagonal matrices.
Exactly the same result holds also for \(B_{2}(\bb{C}[z])\): namely, the latter group is isomorphic to the
semidirect product of its normal subgroup \(U_{2}(\bb{C}[z])\) consisting of matrices of the form \(\smm{1
& 0
\\
p & 1}\) for some \(p\in \bb{C}[z]\) (which is in fact isomorphic to the abelian group \((\bb{C}[z],+)\))
and its subgroup of diagonal matrices, which is again \(D_{2}(\bb{C})\simeq \bb{C}^{*}\times \bb{C}^{*}\).
It follows that every element of \(B_{2}(\bb{C}[z])\) can be uniquely written as a~product of the form
\(ud\) with \(u\in U_{2}(\bb{C}[z])\) and \(d\in D_{2}(\bb{C})\).
Since
\begin{gather*}
\begin{pmatrix}
\alpha&0
\\
0&\beta
\end{pmatrix}
\begin{pmatrix}
1&0
\\
p&1
\\
\end{pmatrix}
\begin{pmatrix}
\alpha^{-1}&0
\\
0&\beta^{-1}
\end{pmatrix}
=
\begin{pmatrix}
1&0
\\
\frac{\beta}{\alpha}p&1
\end{pmatrix},
\end{gather*}
we see that, abstractly, the action of \(\bb{C}^{*}\times \bb{C}^{*}\) on \((\bb{C}[z],+)\) def\/ining the
above semidirect product structure is given by
\begin{gather}
\label{eq:smdp-act}
(\alpha,\beta).p=\frac{\beta}{\alpha}p.
\end{gather}

\section{Proof of the results}

Our strategy to def\/ine the morphism~\eqref{eq:map} is the following.
First, we identify the action on \(\mathcal{C}_{n,2}\) of some strictly op-triangular automorphism in
\(\TAut(A;c)\) with the action of a~unipotent matrix of the form \(\smm{1 & 0
\\ 
p & 1}\) via the f\/low induced by some particular Hamiltonians of the Gibbons--Hermsen system (using
Theorem~\ref{teo:flows} below).
In this way we obtain an embedding of the group \(U_{2}(\bb{C}[z])\) in \(\TAut(A;c)\) which is easily
extended to the whole subgroup \(B_{2}(\bb{C}[z])\).
The subgroup \(\GL_{2}(\bb{C})\subset \GL_{2}(\bb{C}[z])\) consisting of invertible scalar matrices can
also be embedded in \(\TAut(A;c)\) using af\/f\/ine automorphisms acting only on the subspace spanned by
\(x\), \(x^{*}\), \(y\) and \(y^{*}\).
By Theorem~\ref{teo:nagao} these embeddings extend to a~unique morphism of groups
\(\app{k}{\GL_{2}(\bb{C}[z])}{\TAut(A;c)}\).
Finally we use \(k\) to induce the desired morphism \(\app{i}{\Gamma^{\mathrm{alg}}}{\PTAut(A;c)}\).

An automorphism \(\varphi\in \Aut A\) acts on \(\Rep(\ol{Q}_{\mathrm{BP}},(n,1))\), and hence on
\(\mathcal{C}_{n,2}\), in the following way.
For every arrow \(r\) in \(\ol{Q}_{\mathrm{BP}}\), \(\varphi(r)\) is a~non-commutative polynomial in the
arrows of \(\ol{Q}_{\mathrm{BP}}\); in particular we can evaluate it on a~point \(p =
(A,B,X_{1},X_{2},Y_{1},Y_{2})\) (by mapping each arrow to its matrix representation), and this gives
a~matrix \(\varphi(r)(p)\).
Then \(\varphi\) sends \(p\) to the point
\begin{gather}
\label{eq:act-aut}
(\varphi(a)(p),\varphi\big(a^{*}\big)(p),\varphi(x)(p),\varphi(y^{*})(p),\varphi(y)(p),\varphi(x^{*})(p)).
\end{gather}

\begin{example}
If \(\varphi\) is the strictly triangular automorphism \(\Lambda(f)\) with \(f = aab\) then
\begin{gather*}
\frac{\partial f}{\partial a}=ab+ba \qquad \text{and}\qquad \frac{\partial f}{\partial b}=aa,
\end{gather*}
so that \(\varphi\) acts by the formula
\begin{gather*}
\varphi.(A,B,X_{1},X_{2},Y_{1},Y_{2})=\big(A,B+AX_{1}Y_{1}+X_{1}Y_{1}A,X_{1},X_{2}+A^{2}X_{1},Y_{1},Y_{2}
+Y_{1}A^{2}\big).
\end{gather*}
\end{example}

We are now going to prove a~result that enables us to identify the action of some Hamiltonians functions on
\(\mathcal{C}_{n,2}\) with the action of some triangular (or op-triangular) automorphisms in \(\TAut(A;c)\).
This correspondence will be established in much more generality than what is needed in the sequel, since it
may be of independent interest.

Let us def\/ine a~linear map \(H\) from the complex vector space \(L_{2}\) def\/ined in~\eqref{eq:def-L2}
to the ring of regular functions on \(\mathcal{C}_{n,2}\) as follows.
Any element of \(L_{2}\) can be written as a~linear combination of necklace words \(f =
a^{k_{1}}b^{\ell_{1}}\cdots a^{k_{n}}b^{\ell_{n}}\) with \(n\geq 1\) and \(k_{1}, \dots, k_{n}, \ell_{1},
\dots, \ell_{n}\in \bb{N}\) not all zero.
We set
\begin{gather}
\label{eq:def-H}
H(f)\mathrel{:=}\tr X^{k_{1}}(ve_{12}w)^{\ell_{1}}\cdots X^{k_{n}}(ve_{12}w)^{\ell_{n}}
\end{gather}
(where \(e_{12}\mathrel{:=} \smm{0 & 1
\\ 
0 & 0}\)) and extend this by linearity to the whole of \(L_{2}\).
Similarly, we can def\/ine a~map \(H'\) from \(L_{2}^{\mathrm{op}}\) to the ring of regular functions on
\(\mathcal{C}_{n,2}\) by sending the generic necklace word \(f = a^{*k_{1}}b^{*\ell_{1}}\cdots
a^{*k_{n}}b^{*\ell_{n}}\) in \(a^{*}\) and \(b^{*}\) to
\begin{gather}
\label{eq:def-Hp}
H'(f)\mathrel{:=}\tr Y^{k_{1}}(ve_{21}w)^{\ell_{1}}\cdots Y^{k_{n}}(ve_{21}w)^{\ell_{n}},
\end{gather}
where \(e_{21}\mathrel{:=} \smm{0 & 0
\\
1 & 0}\).

\begin{theorem}
\label{teo:flows}
The action determined by the flow at unit time of the Hamiltonian function \(H(f)\) $($resp.\
\(H'(f))\) on \(\mathcal{C}_{n,2}\) coincides with the action~\eqref{eq:act-aut} of the automorphism
\(\Lambda(-f)\) $($resp.\
\(\Lambda'(-f))\).
\end{theorem}

\begin{proof}
By a~straightforward, if tedious, calculation one can verify that the f\/low of \(H(f)\) is given by
solving the following system of dif\/ferential equations:
\begin{subequations}
\begin{gather}
\label{eq:ev-X}
\dot{X}=0,
\\
\label{eq:ev-Y}
\dot{Y}=-\sum_{j=1}^{n}\sum_{i_{j}=1}^{k_{j}}X^{k_{j}-i_{j}}(ve_{12}w)^{\ell_{j}}X^{k_{j+1}}(ve_{12}
w)^{\ell_{j+1}}\cdots X^{k_{j-1}}(ve_{12}w)^{\ell_{j-1}}X^{i_{j}-1},
\\
\label{eq:ev-V}
\dot{v}=\sum_{j=1}^{n}\sum_{i_{j}=1}^{\ell_{j}}(ve_{12}w)^{\ell_{j}-i_{j}}X^{k_{j+1}}(ve_{12}w)^{\ell_{j+1}
}\cdots X^{k_{j}}(ve_{12}w)^{i_{j}-1}ve_{12},
\\
\label{eq:ev-W}
\dot{w}=-e_{12}w\sum_{j=1}^{n}\sum_{i_{j}=1}^{\ell_{j}}(ve_{12}w)^{\ell_{j}-i_{j}}X^{k_{j+1}}(ve_{12}
w)^{\ell_{j+1}}\cdots X^{k_{j}}(ve_{12}w)^{i_{j}-1},
\end{gather}
\end{subequations}
where \(j\) is understood as a~cyclic index modulo \(n\), i.e.\
\(k_{n+1} = k_{1}\) and \(\ell_{n+1} = \ell_{1}\).
These equations can be easily integrated.
Indeed, equation~\eqref{eq:ev-X} and ``half'' of equations~\eqref{eq:ev-V} and~\eqref{eq:ev-W} tell us that
\(X\), \(v_{\bullet 1}\) and \(w_{2\bullet}\) are constants; then the time derivatives of \(v_{\bullet
2}\), \(w_{1\bullet}\) and \(Y\) involve only these constants, so that the f\/lows are linear in time.
Thus the non-trivial part of the f\/low is given by
\begin{gather*}
Y(t)=Y-t\sum_{j=1}^{n}\sum_{i_{j}=1}^{k_{j}}X^{k_{j}-i_{j}}(v_{\bullet1}w_{2\bullet})^{\ell_{j}}
\cdots(v_{\bullet1}w_{2\bullet})^{\ell_{j-1}}X^{i_{j}-1},
\\
v_{\bullet2}(t)=v_{\bullet2}+t\sum_{j=1}^{n}\sum_{i_{j}=1}^{\ell_{j}}(v_{\bullet1}w_{2\bullet})^{\ell_{j}
-i_{j}}X^{k_{j+1}}\cdots X^{k_{j}}(v_{\bullet1}w_{2\bullet})^{i_{j}-1}v_{\bullet1},
\\ 
w_{1\bullet}(t)=w_{1\bullet}-t w_{2\bullet}\sum_{j=1}^{n}\sum_{i_{j}=1}^{\ell_{j}}(v_{\bullet1}w_{2\bullet}
)^{\ell_{j}-i_{j}}X^{k_{j+1}}\cdots X^{k_{j}}(v_{\bullet1}w_{2\bullet})^{i_{j}-1}.
\end{gather*} 
Using the map~\eqref{eq:m1} we can see the above as the following f\/low on
\(\Rep(\ol{Q}_{\mathrm{BP}},(n,1))\):
\begin{gather*}
A(t)=A,
\qquad
X_{1}(t)=X_{1},
\qquad
Y_{1}(t)=Y_{1},
\\
B(t)=B-t\sum_{j=1}^{n}\sum_{i_{j}=1}^{k_{j}}A^{k_{j}-i_{j}}(X_{1}Y_{1})^{\ell_{j}}\cdots(X_{1}Y_{1}
)^{\ell_{j-1}}A^{i_{j}-1},
\\
X_{2}(t)=X_{2}-t\sum_{j=1}^{n}\sum_{i_{j}=1}^{\ell_{j}}(X_{1}Y_{1})^{\ell_{j}-i_{j}}A^{k_{j+1}}
\cdots A^{k_{j}}(X_{1}Y_{1})^{i_{j}-1}X_{1},
\\
Y_{2}(t)=Y_{2}-t Y_{1}\sum_{j=1}^{n}\sum_{i_{j}=1}^{\ell_{j}}(X_{1}Y_{1})^{\ell_{j}-i_{j}}A^{k_{j+1}}
\cdots A^{k_{j}}(X_{1}Y_{1})^{i_{j}-1}.
\end{gather*}
Evaluating at \(t=1\) we recover exactly the action of the automorphism \(\Lambda(-f)\), as can be easily
verif\/ied.
A completely analogous calculation shows that the same relationship holds in the ``opposite'' case between
\(H'(f)\) and \(\Lambda'(-f)\).
\end{proof}

\begin{remark}
The Poisson brackets between the Hamiltonians~\eqref{eq:def-H}, \eqref{eq:def-Hp} are easily calculated in
the following manner.
The vector spaces \(L_{2}\) and \(L_{2}^{\mathrm{op}}\) can be seen as subspaces of the path algebra
\(\bb{C}\ol{Q}_{8}\), where \(Q_{8}\) is the quiver with a~single vertex and two loops \(a\) and \(b\) on
it.
Then the four \(n\times n\) matrices \((X, Y, ve_{12}w, ve_{21}w)\) def\/ine a~point in the representation
space \(\Rep(\ol{Q}_{8},(n))\) and the maps \(H\) and \(H'\) are just the restrictions to
\(\bb{C}Q_{8}/[\bb{C}Q_{8}, \bb{C}Q_{8}]\) and \(\bb{C}Q_{8}^{\mathrm{op}}/[\bb{C}Q_{8}^{\mathrm{op}},
\bb{C}Q_{8}^{\mathrm{op}}]\), respectively, of the map
\begin{gather*}
\app{\psi}{\frac{\bb{C}\ol{Q}_{8}}{[\bb{C}\ol{Q}_{8},\bb{C}\ol{Q}_{8}]}}{\bb{C}[\Rep(\ol{Q}_{8}
,(n))]^{\GL_{n}(\bb{C})}}
\end{gather*}
def\/ined by Ginzburg in~\cite{ginz01}.
There it is proved that \(\psi\) is in fact a~Lie algebra morphism, so that the Poisson bracket between
\(H(f_{1})\) and \(H(f_{2})\) (or \(H'(f_{2})\)) is simply the image of the necklace Lie bracket
\([f_{1},f_{2}]\) under \(\psi\).
It follows in particular that all the Hamiltonians in the image of \(H\) Poisson-commute (and similarly for
\(H'\)); however \(\{H(f_{1}),H'(f_{2})\}\neq 0\) in general.
\end{remark}

Notice that the usual Hamiltonians~\eqref{eq:fh} of the Gibbons--Hermsen system can only give a~polynomial
f\/low on \(\mathcal{C}_{n,2}\) when \(\alpha\) is either the identity (in which case \(\tr Y^{k}vw = \tau
\tr Y^{k}\), as a~consequence of the moment map equation \([X,Y]+vw = \tau I\)) or a~nilpotent matrix.
In what follows we will consider in particular the Hamiltonians \(J_{k,e_{21}} = H'(a^{*k}b^{*})\) (but see
Remark~\ref{rem:ot-h} below).
Under the correspondence~\eqref{eq:corr}, such Hamiltonians correspond to matrices of the form
\(z^{k}e_{21}\).
The exponential of a~linear combination of matrices of this kind,
\begin{gather*}
\sum_{k}p_{k}z^{k}e_{21}
\end{gather*}
is the lower unitriangular matrix \(\smm{1 & 0
\\
p & 1}\), where \(p\) is the polynomial with coef\/f\/icients \(p_{k}\).
Theo\-rem~\ref{teo:flows} then suggests that these elements of \(\GL_{2}(\bb{C}[z])\) should correspond to
the op-triangular automorphisms \(\Lambda'(-p(a^{*})b^{*})\) in \(\TAut(A;c)\).
We are now going to prove Theorem~\ref{teo:result} by building the morphism \(i\) along those lines.

\begin{proof}[Proof of Theorem~\ref{teo:result}]
In view of Theorem~\ref{teo:nagao}, the f\/irst goal is to def\/ine two
morphisms of groups
\begin{gather*}
\app{j_{1}}{\GL_{2}(\bb{C})}{\TAut(A;c)}
\qquad
\text{and} 
\qquad
\app{j_{2}}{B_{2}(\bb{C}[z])}{\TAut(A;c)},
\end{gather*}
that agree on \(B_{2}(\bb{C})\).
We def\/ine \(j_{1}\) by sending \(T\in \GL_{2}(\bb{C})\) to the af\/f\/ine symplectic automorphism
determined by the pair \((I,T)\), where \(I = 0\oplus \id\) is the identity in \(\bb{C}^{2}\rtimes
\SL(2,\bb{C})\).
To def\/ine~\(j_{2}\), notice f\/irst that the subgroup of \(\TAut(A;c)\) consisting of strictly
op-triangular automorphisms of the form \(\Lambda'(-p(a^{*})b^{*})\) for some polynomial \(p\) is
isomorphic to \((\bb{C}[z],+)\).
Moreover, let \(d = \diag(\alpha,\beta)\) be any diagonal matrix in \(\GL_{2}(\bb{C})\); then a~simple
calculation shows that, for every \(p\in \bb{C}[z]\),
\begin{gather*}
(I,d)\circ\Lambda'(-p(a^{*})b^{*})\circ(I,d^{-1})=\Lambda'\left(-\frac{\beta}{\alpha}p(a^{*})b^{*}\right).
\end{gather*}
This is exactly the action~\eqref{eq:smdp-act} def\/ining the semidirect product structure of
\(B_{2}(\bb{C}[z])\), hence we can def\/ine \(j_{2}\) as the unique morphism of groups sending a~lower
unitriangular matrix \(u = \smm{1 & 0
\\
p & 1}\in U_{2}(\bb{C}[z])\) to the automorphism \(\Lambda'(-p(a^{*})b^{*})\) and a~diagonal matrix \(d\in
D_{2}(\bb{C})\) to the af\/f\/ine automorphism \((I,d)\).

With these def\/initions it is immediate to verify that \(j_{2}\) agrees with \(j_{1}\) on
\(B_{2}(\bb{C})\); then by the universal property of amalgamated free products there exists a~unique
morphism of groups \(\app{k}{\GL_{2}(\bb{C}[z])}{\TAut(A;c)}\) whose restriction to \(\GL_{2}(\bb{C})\),
resp.
\(B_{2}(\bb{C}[z])\), coincides with \(j_{1}\), resp.~\(j_{2}\).
It is clear that~\(k\) descends to a~well-def\/ined morphism of groups
\(\app{\tilde{k}}{\PGL_{2}(\bb{C}[z])}{\PTAut(A;c)}\).
We extend \(\tilde{k}\) to \(\Gamma^{\mathrm{alg}} = \Gamma^{\mathrm{alg}}_{\mathrm{sc}} \times
\PGL_{2}(\bb{C}[z])\) as follows.
Let us def\/ine a~morphism of groups \(\app{j_{3}}{\Gamma^{\mathrm{alg}}_{\mathrm{sc}}}{\TAut(A;c)}\) by
sending the generic scalar matrix \(e^{p}I\) (where \(p\in z\bb{C}[z]\)) to the automorphism
\(\Lambda'(-p(a^{*}))\), whose only nontrivial action on the generators is
\begin{gather}
\label{eq:sc-au}
a\mapsto a-\frac{\partial}{\partial a^{*}}p(a^{*}).
\end{gather}
It is easy to verify that such an automorphism commutes with every element in the image of~\(\tilde{k}\),
since it commutes with both elements in the image of \(\Lambda'\) and af\/f\/ine automorphisms of the form
\((I,T)\).
Thus we can def\/ine~\(i\) by mapping the generic element \(\mathrm{e}^{p}M\in \Gamma^{\mathrm{alg}}\) to
the product \(j_{3}(\mathrm{e}^{p}) \tilde{k}(M)\) in \(\PTAut(A;c)\).

Now consider the subgroup \(\mathcal{P}\) of \(\PTAut(A;c)\) generated by the image of \(i\) and the
af\/f\/ine symplectic automorphism \(\mathcal{F}\) def\/ined by~\eqref{eq:def-fft}.
Clearly, \(\mathcal{P}\) acts on \(\mathcal{C}_{n,2}\) by restriction of the action of \(\PTAut(A;c)\).
Recall from~\cite{bp08} that the strategy to prove the transitivity of the latter action is f\/irst to move
every point of \(\mathcal{C}_{n,2}\) into the submanifold
\begin{gather*}
M_{n}\mathrel{:=}\set{(X,Y,v,w)\in\mathcal{C}_{n,2}|v_{\bullet2}=0,w_{2\bullet}=0}
\end{gather*}
(isomorphic to the Calogero--Moser space), and then use the fact that \(\TAut(A;c)\) contains a~copy of the
group \(G_{\mathrm{CM}}\) of symplectic automorphisms of the quiver \(\ol{Q}_{\circ}\) which itself acts
transitively on this submanifold.
As recalled in the introduction, the copy of \(G_{\mathrm{CM}}\) inside \(\TAut(A;c)\) is generated exactly
by the automorphisms of the form~\eqref{eq:sc-au} for some \(p\in z\bb{C}[z]\) together with the single
af\/f\/ine symplectic automorphism \((\smm{0 & -1
\\
1 & 0},I)\).
All of them belong to \(\mathcal{P}\) (the latter being simply the composition of \(\mathcal{F}\) with the
image under \(j_{1}\) of \(\smm{0 & -1
\\
1 & 0}\)), so the only problem is again to move every point of \(\mathcal{C}_{n,2}\) into \(M_{n}\) using
an element of \(\mathcal{P}\).

Now take a~point \(p = (X,Y,\smm{v_{\bullet 1} & v_{\bullet 2}}, \smm{w_{1\bullet}
\\
w_{2\bullet}})\in \mathcal{R}_{n,2}\) for which \(X\) is not regular semisimple; then \(Y\) must be
(otherwise \(p\notin \mathcal{R}_{n,2}\)) and the automorphism \(\mathcal{F}^{-1}\) sends \(p\) to the
point \((Y,-X,\smm{v_{\bullet 2} & v_{\bullet 1}}, \smm{w_{2\bullet}
\\
w_{1\bullet}})\) whose f\/irst entry is regular semisimple.
Hence it is enough to prove that points \((X,Y,v,w)\) for which \(X\) is regular semisimple can be taken
into \(M_{n}\) by the group \(\mathcal{P}\).
But Lemma 8.4 in~\cite{bp08} says exactly that one can map such a~point to \(M_{n}\) using only triangular
automorphisms of the form \(\Lambda(-p(a)b)\) for some polynomial \(p\) and af\/f\/ine ones of the form
\((I,T)\) for some \(T\in \PGL_{2}(\bb{C})\).
Using the relation~\eqref{eq:t-opt} we see that all of these automorphisms belong to \(\mathcal{P}\), hence
the result follows.
\end{proof}
\begin{remark}
Using the fact that \(j_{1}\) and \(j_{2}\) are injective and their image is disjoint it is possible to
show that the map \(k\) def\/ined in the above proof is injective on the reduced words in
\(\GL_{2}(\bb{C}[z])\) of length at most 4.
However, we could not prove that \(k\) is injective in general.
\end{remark}
\begin{remark}
\label{rem:ot-h}
The f\/low determined by a~linear combination of the Hamiltonians \(J_{k,e_{12}}\) should correspond to an
\emph{upper} unitriangular matrix in \(\PGL_{2}(\bb{C}[z])\).
Using the equality
\begin{gather*}
\begin{pmatrix}
1&p
\\
0&1
\end{pmatrix}
=
\begin{pmatrix}
0&1
\\
1&0
\end{pmatrix}
\begin{pmatrix}
1&0
\\
p&1
\end{pmatrix}
\begin{pmatrix}
0&1
\\
1&0
\end{pmatrix}
\end{gather*}
one can see that the map \(i\) def\/ined above sends such matrices to tame symplectic automorphisms of the
form
\begin{gather*}
a\mapsto a+\frac{\partial}{\partial a^{*}}(-p(a^{*})b^{*}),
\qquad
x^{*}\mapsto x^{*}-yp(a^{*}),
\qquad
y^{*}\mapsto y^{*}-p(a^{*})x,
\end{gather*}
which are neither strictly triangular nor strictly op-triangular.
This is due to the fact that the Hamiltonians \(J_{k,e_{12}} = \tr Y^{k}ve_{12}w\) belong neither to the
image of \(H\) nor to that of \(H'\).
\end{remark}

\section{Conclusions and outlook}\label{concl}

We would like to conclude by trying to put into a~wider perspective the results obtained in the present
work.
Let us start by reviewing the situation in the Calogero--Moser case, i.e.\
for~\(r=1\).
In~\cite{bw00} it is proved that there exists a~bijective map
\begin{gather}
\label{eq:bwmap}
\app{\beta_{1}}{\bigsqcup_{n\in\bb{N}}\mathcal{C}_{n,1}}{\mathcal{M}_{1}},
\end{gather}
where \(\mathcal{M}_{1}\) is the set of isomorphism classes of nonzero right ideals in the f\/irst Weyl
algebra \(A_{1}\).
Moreover, the group \(G_{\mathrm{CM}}\) of automorphisms of~\(A_{1}\) acts on both the varieties
\(\mathcal{C}_{n,1}\) (as shown in the introduction) and on \(\mathcal{M}_{1}\) (by its natural action);
the map~\eqref{eq:bwmap} intertwines these two actions, and in fact it can be characterized as the
\emph{unique} map doing so~\cite{wils10}.

Consider now the case \(r>1\).
Let \(B_{1}\) denote the localization of the Weyl algebra \(A_{1}\) with respect to nonzero polynomials.
In the paper~\cite{bgk02} the following correspondence between the manifolds \(\mathcal{C}_{n,r}\) and
a~certain kind of right sub-\(A_{1}\)-modules in \(B_{1}^{r} = B_{1}\times \dots \times B_{1}\) (\(r\)
times) is def\/ined.

Call a~sub-\(A_{1}\)-module \(M\) in \(B_{1}^{r}\) \emph{fat} if there exists a~polynomial \(p\in
\bb{C}[z]\) such that
\begin{gather*}
pA_{1}^{r}\subseteq M\subseteq p^{-1}A_{1}^{r}
\end{gather*}
and denote by \(\Gr^{\mathsf{D}}(r)\) the set of fat sub-\(A_{1}\)-modules in \(B_{1}^{r}\).
For every \(M\in \Gr^{\mathsf{D}}(r)\), let \(\sigma(M)\) be the linear subspace in \(\bb{C}(z)^{r}\)
consisting of all the leading coef\/f\/icients of the operators in \(M\).
Finally, def\/ine \(\mathcal{M}_{r}\mathrel{:=} \sigma^{-1}(\bb{C}[z]^{r})\).
Then Baranovsky, Ginzburg and Kuznetsov prove that for each \(r\in \bb{N}\) there is a~bijection
\begin{gather*}
\app{\beta_{r}}{\bigsqcup_{n\in\bb{N}}\mathcal{C}_{n,r}}{\mathcal{M}_{r}}.
\end{gather*}
This map should play the same r\^ole of the map~\eqref{eq:bwmap} in the case \(r=1\), hence one could hope
that \(\beta_{r}\) is also uniquely determined by some equivariance property.
Unfortunately, when \(r>1\) it is not clear which group should take the place of \(G_{\mathrm{CM}}\simeq
\Aut A_{1}\).

A possible candidate is the group of automorphisms of the matrix algebra \(\Mat_{r,r}(A_{1})\), which
reduces to \(\Aut A_{1}\) when \(r=1\).
In this regard, the notes~\cite{wils09} provide the following intriguing argument.
Denote with \(\Gamma(r)\) the group of holomorphic maps \(\bb{C}\to \GL_{r}(\bb{C})\), and consider the
following subgroup of \(\Gamma(r)\), which represents the obvious generalization of the
group~\eqref{eq:def-Galg} to the case \(r\geq 2\):
\begin{gather*}
\Gamma^{\mathrm{alg}}(r)\mathrel{:=}\Gamma^{\mathrm{alg}}_{\text{sc}}(r)\times\PGL_{r}(\bb{C}[z]),
\end{gather*}
where again \(\Gamma^{\mathrm{alg}}_{\mathrm{sc}}(r)\) consists of maps of the form \(\mathrm{e}^{p}I_{r}\)
for some \(p\in z\bb{C}[z]\).
Now let \(A_{1}^{an}\supset A_{1}\) stand for the algebra of dif\/ferential operators on \(\bb{C}\) with
\emph{entire} coef\/f\/icients.
For every \(\gamma\in \Gamma(r)\), the map \(D\mapsto \gamma D\gamma^{-1}\) is an automorphism of the
algebra \(\Mat_{r,r}(A_{1}^{an})\), and Wilson proves in~\cite{wils09} that \(\Gamma^{\mathrm{alg}}(r)\) is
exactly the subgroup of \(\Gamma(r)\) that preserves the subalgebra \(\Mat_{r,r}(A_{1})\).
Hence \(\Gamma^{\mathrm{alg}}(r)\) can actually be seen as a~subgroup of \(\Aut \Mat_{r,r}(A_{1})\).

It turns out that \(\Aut \Mat_{r,r}(A_{1})\) is the semidirect product of the subgroup of \emph{inner}
automorphisms of \(\Mat_{r,r}(A_{1})\) and a~copy of \(\Aut A_{1}\), acting separately on each matrix entry.
In other words, every automorphism of \(\Mat_{r,r}(A_{1})\) has the form
\begin{gather*}
D\mapsto T\sigma(D)T^{-1}
\end{gather*}
for some \(T\in \GL_{r}(A_{1})\) and \(\sigma \in \Aut A_{1}\).
Such an automorphism belongs to \(\Gamma^{\mathrm{alg}}(r)\) exactly when~\(T\) is a~matrix of polynomials
and \(\sigma\) acts as \(\theta\mapsto \mathrm{e}^{p}\theta \mathrm{e}^{-p}\) (i.e.\
it belongs to the family \(\Phi_{p}\) def\/ined by the equations~\eqref{eq:def-cmflows}).
Clearly, the subgroup of \(\Aut \Mat_{r,r}(A_{1})\) isomorphic to \(\Aut A_{1}\) acts on right
sub-\(A_{1}\)-modules of \(B_{1}^{r}\) (by \(\sigma.M\mathrel{:=} \sigma(M)\), the same prescription
working in the \(r=1\) case) and preserve fatness.
However it is not clear to us, at least at the moment, if this def\/inition can be extended to give an
action of \(\Gamma^{\mathrm{alg}}(r)\) (or the whole group \(\Aut \Mat_{r,r}(A_{1})\)) on fat submodules.

Finally, as a~referee pointed out to us, it appears that much of the results of the paper~\cite{bp08} can
be generalized to the case \(r>2\); it would be interesting to see if the constructions in this paper can
also be generalized to higher values of \(r\).
We hope to address these problems in a~future publication.

\subsection*{Acknowledgements}

The authors would like to thank Claudio Bartocci, Yuri Berest, Roger Bielawski, Ugo Bruzzo, Benoit Dherin,
Letterio Gatto, Victor Ginzburg, Hiraku Nakajima, George Wilson and the anonymous referees for some useful
comments about a~previous version of this manuscript.
Both authors are grateful to FAPESP for supporting the present work with the grants 2010/19201-8 (I.M.) and
2011/09782-6 (A.T.).

\pdfbookmark[1]{References}{ref}
\LastPageEnding

\end{document}